\documentclass{emulateapj}
\newcommand{\ga} {\gtrsim}
\newcommand{\la} {\lesssim}
\lefthead{GOULD, GAUDI, \& HAN} \righthead{RESOLVING THE MICROLENSING MASS DEGENERACY}
\def\kms{{\rm km\,s^{-1}}}

\def\au{{\rm AU}}

\def\ph{{\rm ph}}

\def\e{{\rm E}}
\def\rel{{\rm rel}}
\def\sat{{\rm sat}}
\begin{document}

\journalinfo{The Astrophysical Journal, 591, L000-000, 2003 July 1}

\title{Resolving the Microlens Mass Degeneracy for Earth-Mass Planets}

\author{Andrew Gould\altaffilmark{1,2}, B.\ Scott Gaudi\altaffilmark{2,3} and Cheongho Han\altaffilmark{4}}

\altaffiltext{1}{Ohio State University, Department of Astronomy, Columbus, OH 43210}
\altaffiltext{2}{Institute for Advanced Study, Princeton, NJ 08540}
\altaffiltext{3}{Hubble Fellow}
\altaffiltext{4}{Department of Physics, Institute for Basic Science Research,
Chungbuk National University, Chongju 361-763, Korea}

\email{gould@astronomy.ohio-state.edu}

\begin{abstract}

Of all planet-finding techniques, microlensing is potentially the most 
sensitive to Earth-mass planets.  However, microlensing lightcurves generically
yield only the planet-star mass ratio: the mass itself is uncertain to
a factor of a few.  To determine the planet mass, one must measure both
the ``microlens parallax'' $\tilde r_\e$ and source-lens relative proper motion
$\mu_\rel$.  
Here we present a new method to measure microlens masses for terrestrial
planets.  We show that, with only a modest adjustment to the proposed orbit
of the dedicated satellite that finds the events, and combined 
with observations from a ground-based observing program, the planet mass 
can be measured routinely.  The dedicated satellite that finds the
events will automatically measure the proper motion and one projection
of the ``vector microlens parallax'' ($\tilde r_\e,\phi$).  
If the satellite is placed in an L2 orbit,
or a highly elliptical orbit around the Earth, the Earth-satellite baseline
is sufficient to measure a second projection of the vector microlens 
parallax from the difference in the lightcurves as seen from the Earth
and the satellite as the source passes over the caustic structure induced by the planet.  
This completes the mass measurement.

\end{abstract}

\section{Introduction
\label{sec:intro}}

Among all proposed methods to search for extra-solar planets, only microlensing
has the property that the intrinsic amplitude of the planetary signature remains
constant as the planet mass decreases.  Hence, with the notable exception of
pulsar timing \citep{wolz}, microlensing can in principle probe to
lower masses than any other technique.   A microlensing space mission that
was of similar scale to the transit missions
{\it Kepler}\footnote{http://www.kepler.arc.nasa.gov/} and
{\it Eddington}\footnote{http://sci.esa.int/home/eddington/index.cfm}
or to the astrometry satellite
{\it Space Interferometry 
Mission (SIM)}\footnote{http://planetquest.jpl.nasa.gov/SIM/sim\_index.html}
would be sensitive to Mars-mass companions \citep{br2}, a decade or
two below these other techniques.  Furthermore, any microlensing detections 
of terrestrial planets are expected to be at significantly higher 
signal-to-noise ratio (S/N), and thus will be more robust to unforeseen 
systematic errors.   Hence, microlensing can potentially play a major role 
in determining the frequency of terrestrial planets around main-sequence 
stars.  An accurate assessment of this frequency is a key requirement for 
the design of the {\it Terrestrial Planet
Finder}\footnote{http://planetquest.jpl.nasa.gov/TPF/tpf\_index.html},
which will ultimately take images and spectra of such planets.

Unfortunately, while microlensing can detect planets of very low mass,
there has not seemed to be any way to measure the masses of those
planets to better than a factor of a few: although microlensing light curves automatically yield
the planet-star mass ratio $q=m_p/M$, the stellar mass itself is unknown
due to the classic microlensing degeneracy.  This degeneracy arises from the fact that among the three 
microlensing ``observables'', the Einstein timescale $t_\e$, 
the angular Einstein radius $\theta_\e$, and
the projected Einstein radius $\tilde r_\e$, 
only $t_\e$ is routinely extractable from the microlensing
light curve.  These three observables are related to the three underlying
physical parameters, $M$, $\pi_\rel$, and $\mu_\rel$, by
\begin{equation}
t_\e = {\theta_\e\over\mu_\rel},\quad
\theta_\e = \sqrt{4GM\pi_\rel\over c^2\,\au},\quad
\tilde r_\e = \sqrt{4GM\,\au\over c^2\,\pi_\rel}.
\label{eqn:parms}
\end{equation}
Here, $\pi_\rel$ and $\mu_\rel$ are the source-lens relative parallax and
proper motion.  To determine the mass would require measurement of the
other two observables: $M = (c^2/4G)\tilde r_\e \theta_\e$.  To date,
$\tilde r_\e$ and $\theta_\e$ have been measured for only about a dozen events
each out of the more than 1000 so far discovered, and only for one event have
both been measured, thus yielding the mass (\citealt{eros2k5} and references
therein).  

Although several methods to partially break the microlensing mass 
degeneracy in special instances have been proposed \citep{gg,ratten2002,br2},
so far there has only been one idea to do so for 
a large, representative ensemble of events.  \citet{gs} showed that by 
combining observations
from the ground and the solar-orbiting {\it SIM}, one could measure $\theta_\e$
astrometrically and $\tilde r_\e$ photometrically and so routinely measure
the mass.  Unfortunately, this technique cannot be applied to 
terrestrial-planet microlensing events, even in principle.  Terrestrial
planets can only be detected in events of main-sequence source stars.
For giant sources, the planetary microlensing pattern would be much smaller
than the source and so would be undetectable \citep{br1}.
Because of its small aperture,
{\it SIM} cannot observe Galactic-bulge main-sequence stars to the 
required precision.

Here we present a new method to measure microlens masses for terrestrial
planets.  The method requires only a modest adjustment to the orbit of
a microlensing planet-finder satellite and combining its observations
with a ground-based observing program.

\section{Microlensing Parameters from a Single Observer
\label{sec:inventory}}

To understand how $\tilde r_\e$ and $\theta_\e$ can both be measured for
terrestrial planets,
one should first take a careful inventory of what parameters are
automatically measured from planetary microlensing events detected from a 
planet-finder satellite.  

From the width, height and peak time of the underlying event due to
the primary, one obtains the three standard microlensing parameters
$t_\e$, $u_0$, and $t_0$, where the latter two are the dimensionless
impact parameter and time of maximum \citep{pac86}.  In addition to
these usual three parameters, one additional parameter of the primary
event can also be routinely measured: the parallax asymmetry
$\gamma$. The Earth's acceleration during the event induces parallax
effects on the lightcurve.  If the event lasts a substantial fraction
of a year, then these effects can be used to measure both components
of the vector microlens parallax \citep{gould92}.  
However, for more typical short events, the effect reduces to an 
asymmetry in the light curve \citep{gmb}.  This parallax asymmetry is
in effect a projection of the full vector parallax, and as such is
described by a single parameter, $\gamma$,
\begin{equation}
\gamma \equiv a_\oplus {t_\e\over \tilde v}\cos\psi\cos\phi,
\quad \tilde v \equiv {\tilde r_\e\over t_\e},
\quad a_\oplus\equiv {4\pi^2\,\au\over \rm yr^2}.
\label{eqn:gammadef}
\end{equation}
Here $\au\cos\psi$ is the length of the Earth-Sun separation projected onto
the plane of the sky, $\phi$ is the angle between the source trajectory
and this projected separation, and $\tilde v$ is the source-lens relative
speed projected onto the observer plane.   For the typical timescales and
projected velocities of events toward the Galactic bulge, $t_\e \sim 20~{\rm days}$ and
${\tilde v}\sim 800~\kms$, the parallax asymmetries would appear
to be unmeasurable small,   $\gamma \la 10^{-2}$. 
However, high-cadence 
$(f\sim 144\,\rm day^{-1})$, high-precision ($\sigma_\ph \sim 0.01\,\rm mag$)
continuous photometric monitoring is required to detect terrestrial
planets in the first place \citep{br2}.   As a by-product of such photometry,
it should be possible to measure such small parallax asymmetries.
\citet{gould98} showed that $\gamma$ 
can be measured with S/N 
\begin{equation}
{|\gamma|\over \sigma_\gamma} = 12
\biggl({\sigma_\ph\over 0.01}\biggr)^{-1}
\biggl({f\over 144\,\rm day^{-1}}\biggr)^{-1}
\biggl[{S(u_0)\over 3}\biggr]^{-1}
\end{equation}
$$
\times \biggl({\tilde v\over 800\,\kms}\biggr)^{-1}
\biggl({t_\e\over 20\,\rm days}\biggr)^{3/2}
{|\cos\psi\cos\phi| \over 0.5},
\label{eqn:gammasn}
$$
where for observation streams beginning and ending well beyond the
event, $S$ varies monotonically from $S(0)=2.1$ to $S(0.7)=4.4$.
Thus except near June 21 when $\cos\psi\sim 0.1$ for observations
toward the Galactic bulge, and except for extremely short events,
it should be possible to routinely measure $\gamma$ with good S/N.

For all planetary events, it is generally possible to measure an additional
three parameters.   Planets generally induce a short-duration deviation to an otherwise unperturbed standard
microlensing event.  From the duration, peak time,
and size and shape of the planetary perturbation, one obtains the planet-star
mass ratio $q$, the angle $\alpha$ of the planet-star projected separation
relative to the source trajectory, and the angular planet-star
separation in units of $\theta_\e$ \citep{gl}.

Finally, for terrestrial planets 
it should be possible to routinely measure one additional parameter:
$\rho_*\equiv\theta_*/\theta_\e$, where $\theta_*$ is the source radius.
Since $\theta_*$ can be determined from the dereddened color and magnitude
of the source (see e.g., fig.\ 10 from \citealt{eros2k5}), this would yield
$\theta_\e$.  This ratio can be measured whenever the source passes
over a magnification pattern with structure on scales $\la \rho_*$.
In particular, for planetary events, \citet{gg} find that it can be measured
provided that $\rho_*\ga 0.3q^{1/2}$, which corresponds to,
\begin{equation}
m_p \la 4\,M_\oplus
\biggl({\theta_*\over r_\odot/R_0}\biggr)^{2}
\biggl({\pi_\rel\over 40\mu\rm as}\biggr)^{-1},
\label{eqn:mplimit}
\end{equation}
where $r_\odot/R_0$ is the angular size of a solar-type star at the 
Galactocentric distance.  Note that it is also possible to measure $\rho_*$, and
thus $\theta_\e$, from events due to higher-mass planets if
the source crosses the planetary caustic.

\section{Parameters from Two Observers and Degeneracy Resolution
\label{sec:method}}

From satellite measurements alone (\S~\ref{sec:inventory}), most of the pieces 
are already
in place for terrestrial planet mass measurements.  Since both $\theta_\e$
and one combination of the vector microlens parallax 
$(\tilde r_\e,\phi)$ can already be measured,  all
that is required is a measurement of another combination of 
$(\tilde r_\e,\phi)$.  

It is well known that $\tilde r_\e$ can be measured
by observing an event simultaneously from two telescopes that are significantly
displaced from each other.  Here `significantly' means that the light curve
appears measurably different from the two observatories.  In other words, the magnification
pattern being probed must have structure on a scale that, when projected
to the observer plane, is comparable to the separation of the observers.  For
typical primary lensing events toward the Galactic bulge, the projected scale of 
the magnification structure is ${\tilde r_\e} \sim 8~{\rm AU}$,
 and so the observers must be separated by at least $O({\rm AU})$. 
Thus, by observing the event simultaneously from telescopes on the Earth and
in solar orbit, one could routinely measure $\tilde r_\e$ \citep{refsdal,gould95}.

Therefore, at first sight, the solution 
appears simple: just put the microlensing satellite in orbit around the Sun and
carry out simultaneous observations from the ground.  Unfortunately,
the huge data stream from the continuous monitoring of $O(10^9)$ pixels 
required to detect the planets \citep{br2} make
this impossible unless there are major breakthroughs in satellite telemetry.
The satellite must stay reasonably close to the Earth to transmit these data
efficiently.

However, two factors combine to make feasible microlens parallax measurements from short baselines for events with terrestrial planets.
First, one component of the vector microlens parallax is already
measured for these  events because of the extremely high overall
S/N required to detect them. See \S~\ref{sec:inventory}.  Second,
the planetary perturbation has structure on
scales that are smaller than the primary Einstein ring by a factor of
$\sim q^{1/2}$.  Therefore, 
two observers need only be separated by of order the scale
of the structures, not the whole Einstein ring \citep{hw,ga,graff}.  For planetary events, the perturbed regions of the 
Einstein ring typically lie on a line along the planet-star axis and
have a width of order $q^{1/2}\tilde r_\e$.   The satellite will cross
this line at a time that differs from the Earth crossing by $\Delta t$.  Figure
\ref{fig:one}a shows the geometry.  
By applying the Law of Sines, one finds
\begin{equation}
\Delta t = (q^{1/2} t_\e) {d_\sat\over q^{1/2} \tilde r_\e}\,
{\sin(\phi+\beta - \alpha)\over \sin\alpha},
\label{eqn:deltat}
\end{equation}
where $d_\sat$ is the distance to the satellite,
$\alpha$ is the known angle between planet-star axis and the
source trajectory, $\beta$ is the known angle between the Earth-Sun
and Earth-satellite axes, both projected on the sky, and $\phi$ is the
(a priori) unknown angle between the Earth-Sun axis and the source trajectory.
Combining equations (\ref{eqn:gammasn}) and (\ref{eqn:deltat}), one obtains an
explicit expression for $\phi$,
\begin{equation}
\tan\phi = {\Delta t\over \gamma}\,
{a_\oplus t_\e \sin\alpha\cos\psi\over d_\sat\cos(\beta-\alpha)} -
\tan(\beta - \alpha),
\label{eqn:tanphi}
\end{equation}
and by means of equation (\ref{eqn:gammasn}) an explicit expression for
$\tilde v$ as well.

The first term on the right-hand side of equation (\ref{eqn:deltat})
is roughly the duration of the perturbation,
the second term is the dimensionless ratio of the Earth-Satellite separation to the width
of the perturbation, while the third term is of order unity.  Hence,
$\Delta t$ can be measured with a fractional precision
$\sigma(\Delta t)/\Delta t \sim (\tilde r_\e q^{1/2}/d_\sat)/(\Delta \chi^2)^{1/2}$,
where $\Delta\chi^2$ is the square of the S/N with which the perturbation
is detected from the weaker observatory (probably the ground).  The proposed
planet detection threshold from space is $\Delta\chi^2 = 160$, but the expected distribution
has a long tail toward larger values, so that half the detections
have $\Delta\chi^2>800$ \citep{br2}.  Thus, $\Delta t$ could be measured with
reasonable precision for a significant fraction of events provided that
the satellite was not more than a few times closer than the size of the
planetary Einstein ring, $q^{1/2}\tilde r_\e$, and that
the ground-based observations were not more than a few times worse than
the satellite observations.  In addition, the separation cannot be more
than a few planetary Einstein radii or the Earth will pass outside
the region of the planetary perturbation.  To target Earth-mass planets,
the separation should therefore be
\begin{equation}
d_\sat \sim \sqrt{4GM_\oplus\,\au\over c^2\,\pi_\rel} = 
0.025\,\au\biggl({\pi_\rel\over 40\,\mu\rm as}\biggr)^{-1/2}.
\label{eqn:dsat}
\end{equation}
A near optimal solution would seem to be to place the satellite in L2
orbit, which lies at $0.01\,\au$ in the anti-Sun direction.  However,
while data transmission is $10^4$ times more efficient from L2 than
from an AU, that still might not be efficient enough.

A plausible alternative approach would then be to put the satellite in a highly
elliptical orbit with period $P\sim 1\,$month.  It would spend the majority of
its time near $2a\sim 0.005\,\au$, adequate for Earth-mass and lighter planets.
During the brief perigee each month it could focus on highly efficient
data transmission.  Because of this large semi-major axis, the orbit would
have to be well out of the ecliptic to avoid gravitational encounters with
the Moon, but not so far out that the orbit destablized and crashed into
the Earth.  In fact, it might be difficult to find such long-term stable
orbits, but the satellite could be ejected into solar orbit at the end
of its mission with a boost at perigee of only 
$\Delta v\sim 100\,\rm m\,s^{-1}$, thereby evading the requirement for
long-term stability.

One potential concern is that if the satellite is anywhere in the ecliptic
(including L2), then $\beta=0^\circ$ 
(or $180^\circ$).  Microlensing is most sensitive to planets close to the
peak of the event.  At the peak itself, $\alpha=90^\circ$.  Therefore, near
the peak both terms in equation (\ref{eqn:tanphi}) would be very large, which
would in effect magnify the observational errors.  However, we find from
simulations that the enhanced sensitivity at $\alpha=90^\circ$ does not 
imply a tight clustering of events at this value.  Rather the distribution
is extremely broad, so there is only a marginal cost to having the
satellite in the ecliptic.

\section{Discussion
\label{sec:discuss}}

For typical relatively short events, the parallax asymmetry is quite weak
and is only detectable because of the satellite's high cadence and S/N.
Thus, one must worry about systematic effects.  \cite{gould98} identified 
three such effects not specific to terrestrial observers:
variable sources, binary sources, 
and binary lenses.  Because of the long high-quality data stream, the source
can easily be checked for low levels of variability.  While there may be
occasional stars that vary over a few months but not otherwise over several
years, the fraction of such stars is not likely to be large and can be
measured from the prodigious supply of data on ``stable'' stars.  
A binary companion to the source star would have to be separated
 by 2 or 3 $\theta_\e$ and have a flux ratio of $\sim \gamma \sim 1\%$ to 
reproduce the magnitude and shape of a parallax asymmetry.  
Although additional flux at this level would be evident from a fit to 
the microlensing
event itself, it would not be distinguishable from light from the lens
star.  However, one could check for consistency between the amount of blended
light and the mass and distance to the lens as determined from the parallax asymmetry.  
Further, if the source is really a binary, 
high-resolution spectroscopy could uncover of order $10\,\kms$
radial velocity variations over time.   Binary lenses can also induce
asymmetries.  
There are no studies of the expected rate of these, but for
field stars it is probably of the same order as events with pronounced 
deviations, which is $\sim 5\%$.  However, most stars with planets are
unlikely to have binary companions within a factor 3 or so of the Einstein 
radius because they would render the planetary orbit unstable.
Thus, while caution is certainly warranted in interpreting lightcurve
asymmetries as being due to parallax, systematic effects are unlikely
to dominate the signal.

\begin{figure}[htbp]
\plotone{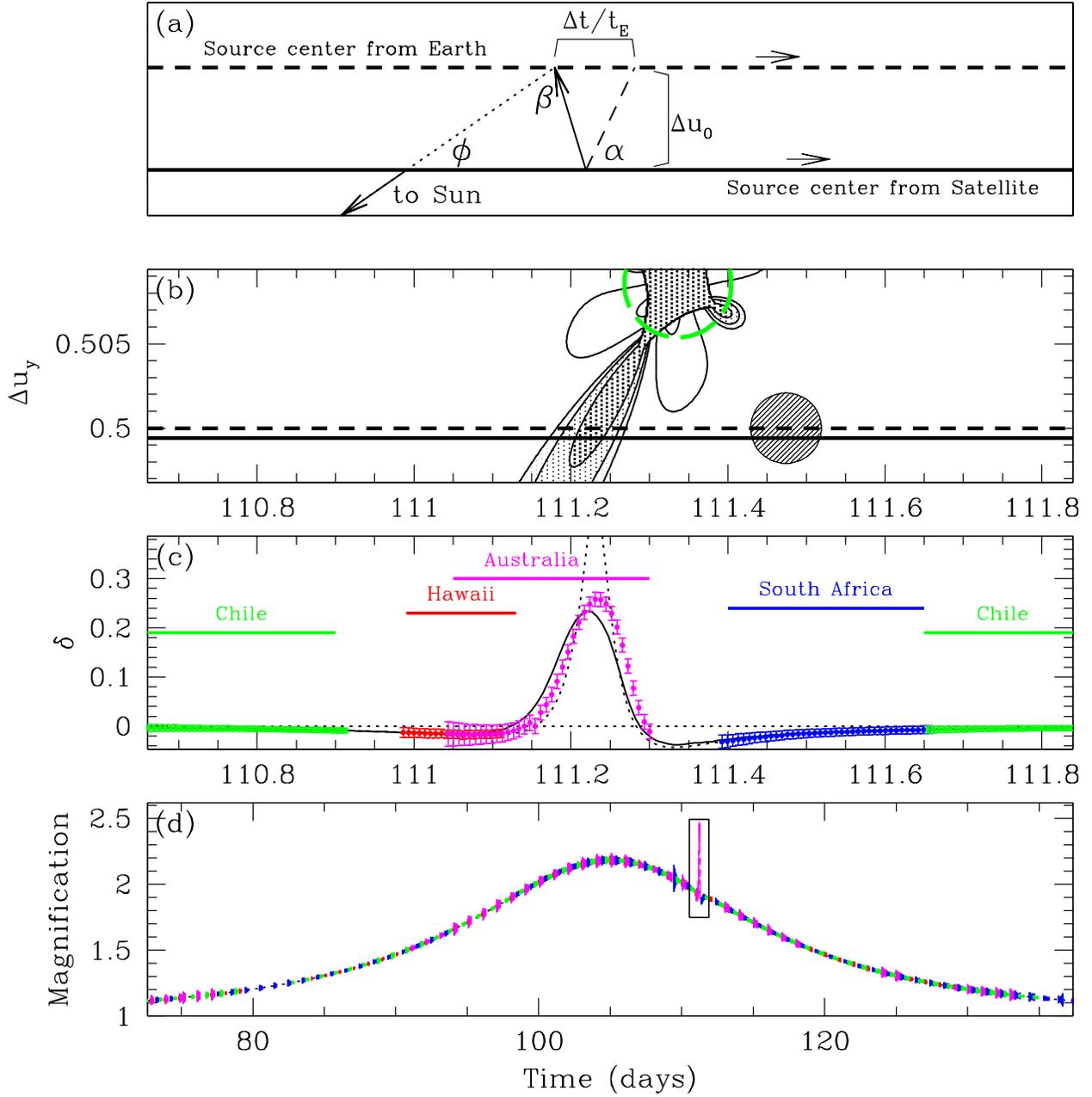}
\caption{\label{fig:one}
(a) Geometry of microlens planetary mass measurement.  All lines
are projected onto the two-dimensional plane of the sky and all distances
are scaled to the projected Einstein radius $\tilde r_\e$.  Source
as seen from the satellite ({\it bold line}) travels horizontally at an 
(a priori unknown) angle $\phi$ relative to the line connecting
the Sun and Earth ({\it dotted line}), which in turn is at a (known) angle
$\beta$ relative to the line connecting the Earth and the satellite.
The star-planet axis ({\it thin dashed line}) lies at a (known) angle 
$\alpha$ relative to the source trajectory.  As seen from the Earth,
the source ({\it bold dashed line}) moves on a parallel trajectory
but displaced by a distance $d_\sat/\tilde r_\e$.  As a result, the
source intersects the perturbation induced by the planet (along the star-planet
axis) at a time later by $\Delta t$, corresponding to a fraction 
$\Delta t/t_\e$ on an Einstein radius. 
(b) Contours of constant fractional deviation $\delta$ from the primary
lensing event calculated for a point source.  
Contour levels are $\delta=\pm 5\%,10\%,25\%$; positive contours
are shaded.  Long-dashed circle shows the planetary Einstein ring radius.  
The shaded
circle shows the size of the source.  Horizontal lines are as in panel (a).
(c) Planetary perturbations from the primary event as seen from the Earth 
({\it points with error bars}) and the satellite ({\it solid curve}), taking
into account of 
the finite size of the source.  The dotted curve is for a point source as seen
from the Earth.  (d) Primary lensing event with planetary perturbation region 
outlined.
In the example shown, $\phi=30^\circ$, $\beta=80^\circ$,
$\alpha=60^\circ$, $\cos\psi=0.93$, $\tilde r_\e=8\,\au$,
$d_\sat=0.0054\,{\rm AU}$, $t_\e=21.58\,$days, $\rho_*=2.1\times
10^{-3}$, $q=10^{-5}$, $\pi_\rel=38\mu{\rm as}$, $\tilde v =630\kms$,
$M=0.3M_\odot$, and $m_p=q M=M_\oplus$.
}\end{figure}

There are two types of checks that can be performed on the mass measurements
derived by our method.  First, in a significant minority planetary events, the
lens can be directly observed \citep{br2}.  The derived mass 
and relative lens-source parallax ($\pi_\rel/\au = \theta_\e/\tilde r_\e$) 
can then be compared to 
the same quantities as determined from multi-color photometry and/or
spectroscopy.  Second, in 
some cases
it will be possible to measure not only the offset parallel to the source-lens
relative motion $\Delta t/t_\e$, 
but also the offset in the orthogonal direction
$\Delta u_0$.  This is because the source will pass over a different part
of the planetary perturbation, which will generally yield a slightly 
different perturbation magnitude (see Fig.~\ref{fig:one}c). 
Measurement of both the difference in the magnitude and time of the 
perturbation then gives the two-dimensional offset in the Einstein ring, 
and thus a measurement of both components of $(\tilde r_\e,\phi)$.   This
effect is typically weaker than the time offset because the
magnification contours as stretched along the planet-star axis, 
and so requires a higher S/N to detect, but in the cases for which it is
detected, the result can be cross checked against the asymmetry measurement.

A microlens planet-finding satellite with parallax capabilities would have
a number of other applications.  First, it would automatically make
precise mass measurements on all caustic-crossing binaries \citep{graff}.
Second, although it would not measure masses for the majority
of larger planets such as gas giants, it would do so for the significant
minority of cases in which the source passed over the planetary caustic.
 From a mathematical point of view, these cases are identical to the
caustic-crossing binaries analyzed by \citet{graff}.  These caustics
are substantially larger than the entire perturbation due to an Earth
mass planet. Hence, if there are equal numbers of Earth-mass and 
Jovian-mass planets, the latter will yield 
the majority of the mass measurements
even though the fraction of mass measurements is higher among the former.

Finally, we have so far not given much attention to the problem of 
organizing the round-the-clock (and so round-the-world) ground-based 
observations that must complement the satellite observations.  Although
easier and cheaper than launching a satellite, the effort
required for this is by no means trivial.  Such a survey would have
tremendous potential in its own right and might be undertaken
independently of a satellite.  We reserve a full discussion of this
idea to a future paper.

\acknowledgments 
Work by AG was supported by JPL contract 1226901. 
Work by SG was supported by NASA through
a Hubble Fellowship grant from the Space Telescope Science Institute,
which is operated by the Association of Universities for Research in
Astronomy, Inc., under NASA contract NAS5-26555. 
Work by CH was supported by the Astrophysical Research Center
for the Structure and Evolution of the Cosmos (ARCSEC) of Korea
Science and Engineering Foundation (KOSEF) through the Science
Research Center (SRC) program.

\end{document}